# A proposal for imaging spectro-polarimetry with a new generation Multichannel Subtractive Double Pass (MSDP) onboard the EST telescope

*Malherbe, J.-M. (1), Sayède, F. (2), Mein P. (3)*
*Emérite LESIA (1), GEPI (2), Honoraire LESIA (3), Observatoire de Paris, 92195 Meudon, France*

*October 5, 2023*

**Abstract**

Imaging spectroscopy is intended to be coupled with adaptive optics (AO) on large telescopes, such as EST, in order to produce high spatial and temporal resolution measurements of velocities and magnetic fields upon a 2D FOV. We propose a Multichannel Subtractive Double Pass (MSDP) incorporated to the EST visible and IR spectrographs, using new generation slicers (56 channels, high spectral resolution) which will benefit of AO and polarimeters. The aim is to produce 56-channel spectra images with the spatial resolution of the AO and reconstitute **cubes** of **instantaneous data (X, Y, lambda)** at high cadence, allowing the study of the plasma dynamics and magnetic fields. The MSDP is compatible with most polarimetric methods (we shall discuss only two of them).

## 1 - The MSDP principle

The Multichannel Subtractive Double Pass is an imaging spectroscopy device which can be incorporated to most grating spectrographs (Mein et al, 2021). The first system was built for the 14 m spectrograph of the Meudon Solar Tower (Mein, 1977). The MSDP principle is described by Figures 1 and 2. A second MSDP was put in the 8 m spectrograph of The Pic du Midi Turret Dome in the eighties (Mein, 1980, Figure 3). The third one was introduced into the 15 m spectrograph of the german Vacuum Tower Telescope (VTT) in Tenerife (Mein, 1991, Figure 4). The most recent MSDP was designed for the 8 m spectrograph of the THEMIS telescope (Mein, 2002). Meanwhile, a collaboration started with polish colleagues, and a new MSDP system joined the large Wroclaw coronagraph in Bialkow. The MSDP can also be operated in polarimetric mode (Figure 5) and is compatible with many polarimeters.

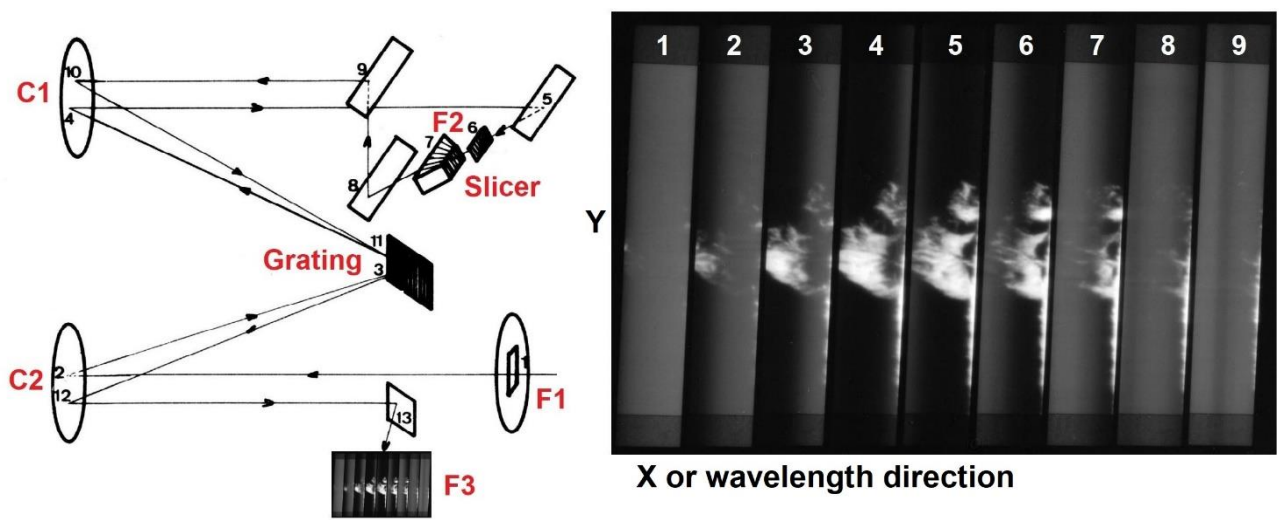

*Figure 1: the MSDP principle. The MSDP is composed of a beam splitter shifter (the slicer 6, 7) and associated flat mirrors (5, 8, 9) located at the F2 focus of a classical grating spectrograph with the collimator*

*(C1) and the chamber (C2). F1 is a rectangular entrance window in the solar image. The first pass (1, 2, 3, 4) forms the spectrum in F2. Several channels are selected by the slicer which re-injects the beams into the spectrograph for a subtractive second pass (10, 11, 12, 13). A multichannel spectra image is formed at F3. Here, the example shows an Hα spectra image with 9 channels. Today, the slicers work with 24 channels. Increasing the channel number allows to improve the spectral resolution and the bandpass of the system to cover the line core and wings over a large 2D FOV.*

The core of the MSDP is a slicer which allows to select several 2D channels along the line profiles; previous slicers used multislit beam-splitters and associate prism beam-shifters. The new generation slicers use cheap and flat (micro) mirrors for beam splitting and shifting, allowing to improve the spectral resolution and the transmitted light. The Solar Line Emission Dopplerometer (SLED, Malherbe *et al*, 2021) uses a compact slicer made of one block of 24 micro-mirrors (the beam splitter) and 24 adjustable secondary mirrors (the beam shifter); this compact MSDP (2 m focal length spectrograph) will be installed on the Zeiss coronagraph of Lomnicky (2640 m) in Slovakia to study the dynamics of the hot corona (forbidden iron lines).

It must be noticed that the raw channels are not monochromatic (Figure 2). If x designates the abscissa along the channels (mm), Δx the slicer step (mm), n the current channel (n varying from 1 to N) and d the spectrograph dispersion (in mm/Å), the wavelength varies in the x-direction of each channel such as:

$$\lambda_n(x) = \lambda_0 + (x/d) + n\,(\Delta x/d)$$

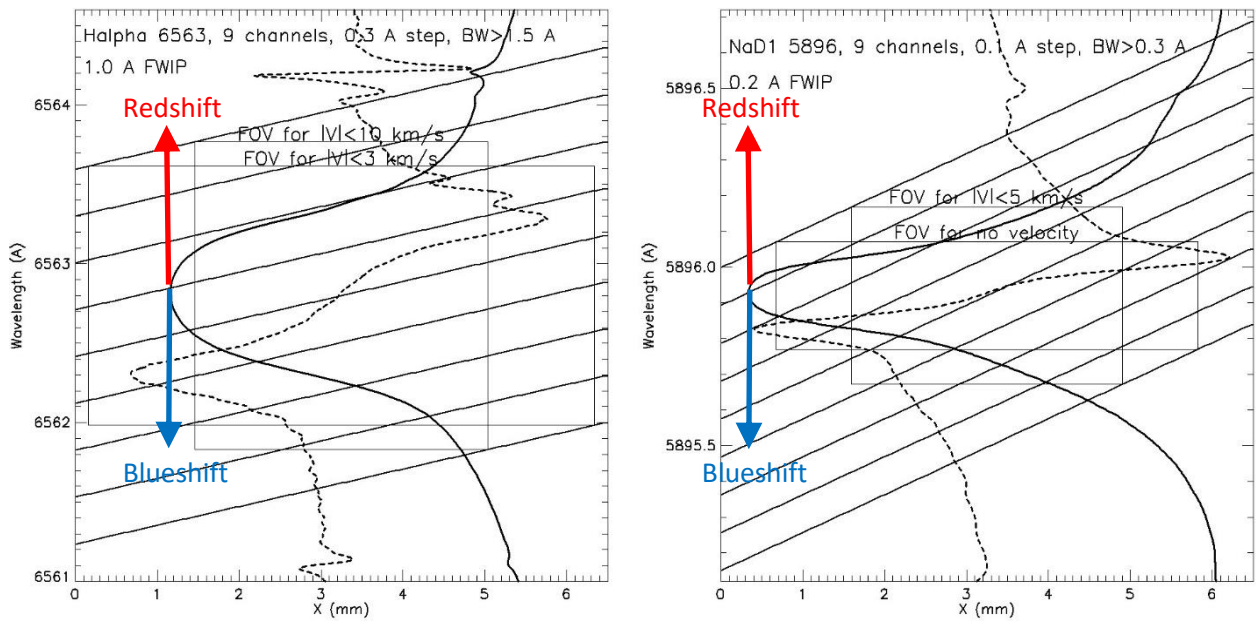

*Figure 2: the MSDP wavelength transmission $\lambda_n(x)$ for the 9 Hα channels (left, 0.3 Å resolution) and the 9 NaD1 channels (right, 0.1 Å resolution) at Meudon Solar Tower. Hα is a broad line (1.0 Å FWIP, for Full Width at Inflexion Points), NaD1 is thinner (0.2 Å FWIP). The line profile at disk centre and its first derivative are displayed. For a given bandpass centred on the line core (here 1.5 Å for Hα, 0.3 Å for NaD1), the graphs show that the measure of large Dopplershifts is possible in a reduced X-direction FOV; of course, it is possible to deal with high velocities with more channels than nine.*

$\lambda_0$ is a reference value. For a given abscissa in the 2D FOV, there is a constant wavelength step (Δx/d) between each channel. However, the sampling wavelengths $\lambda_n$ of the line profile depends on the location in the x-direction : for x = 0 (left of the FOV), sampling values are $\lambda_n = \lambda_0 + n\,(\Delta x/d)$, while for $x = x_m$ (right of the FOV), $\lambda_n = \lambda_0 + (x_m + n\,\Delta x)/d$. For a 8 m focal length spectrograph, the typical dispersion is d ≈ 5 mm/Å (larger in the blue, smaller in the red). The slicer step Δx can be set in the range 0.18 mm (EST project) or less to more than 2 mm (historic Meudon slicers); this value is imposed by the spectral resolution (Δx/d) needed for the lines, which can be in the range 0.03 Å (or less) to 0.3 Å, depending on the spectrograph dispersion (d) and the line width requirements. The total number of channels N lies in the range 7 to 56. The first MSDP had only 7 channels; THEMIS MSDP had 16 channels, SLED has 24, while 56 are forecasted for EST. With a high spectral resolution (Δx/d small), it is necessary to have a large number (N) of channels, otherwise the spectral coverage of the line would not suffice. For EST, Δx = 0.18 mm should provide (for lines around 500 nm wavelength) the spectral resolution (Δx/d) of about 30 mÅ (R = 167000), while N = 56 should guarantee a large spectral coverage of about 1 Å (the maximum possible is N Δx/d = 1.8 Å upon a reduced FOV). For specific needs, a slicer step as low as Δx = 0.10 mm is possible to reach the resolution R = 300000. Table 1 summarizes the capabilities of the existing MSDP. Most use multislit slicers, except the recent instruments which use micromirrors, a new technology allowing to increase the spectral resolution, the number of channels and the luminosity.

| MSDP | step (Δx, mm) | slicer type | channels (N) | translation (mm) | Focal length |
|---|---|---|---|---|---|
| Meudon | 2.5 | multislit | 9 | 9 | 14 m |
| (1977) | 1.0 | multislit | 9 | 9 | |
| | 0.3 | micromirrors | 18 | 3.3 | |
| Pic du Midi | 1.2 | multislit | 11 | 7.5 | 8 m |
| (1980) | 0.6 | multislit | 11 | 7.5 | |
| VTT Izana | 2.5 | multislit | 9 | 9 | 15 m |
| (1990) | 1.2 | multislit | 9 | 9 | |
| | 0.6 | multislit | 11 | 7.5 | |
| (THEMIS | 1.2 | multislit | 9 | 9 | 8 m |
| (2000) | 0.4 | multislit | 16 | 2.5 | |
| BIALKOW | 1.2 | multislit | 9 | 9 | 14.5 m |
| | | | | | |
| SLED (2024) | 0.4 | micromirrors | 24 | 3.3 | 2 m |
| EST (2030) ? | 0.18 | micromirrors | 56 | 6.2 | 7-8 m |
| | 0.27 | micromirrors | 56 | 6.2 | |
| | 0.10 (optional) | micromirrors | 56 | 6.2 | |

*Table 1: optical capabilities of existing and future MSDP*

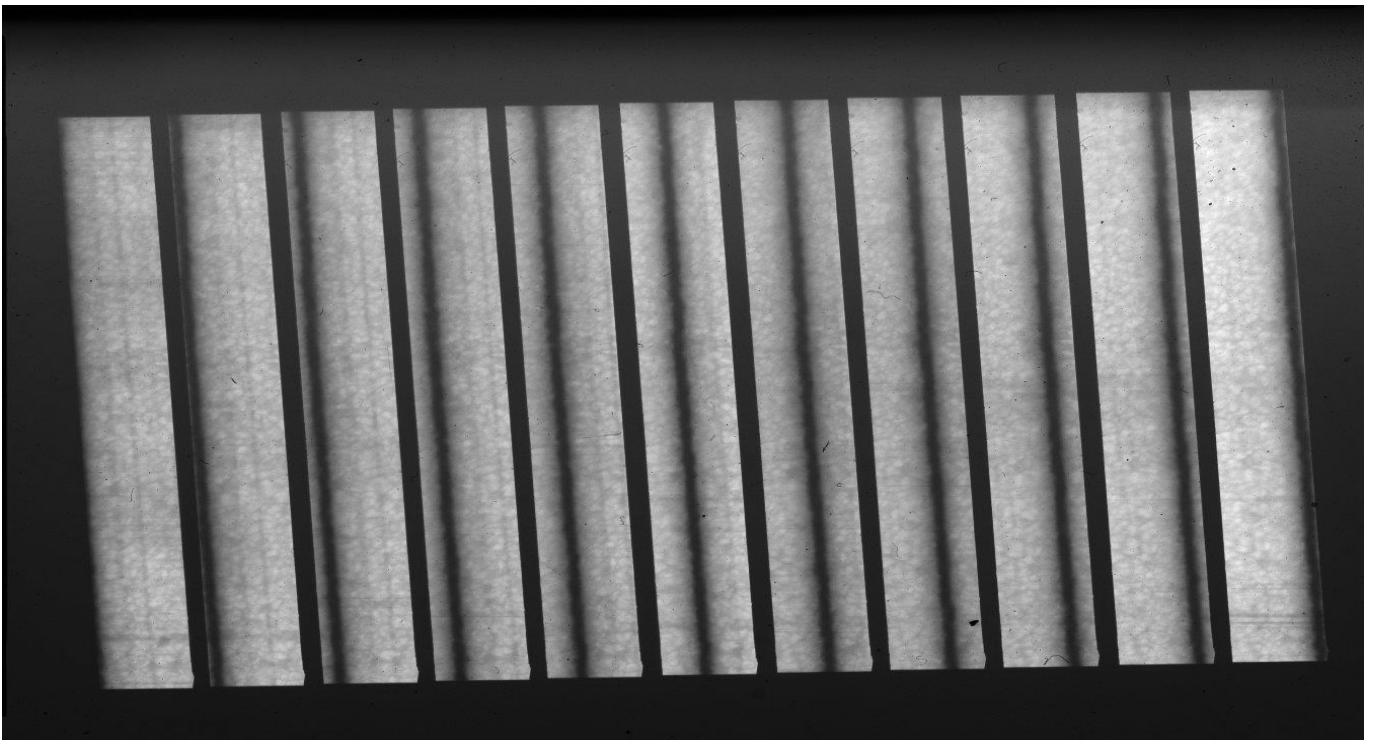

*Figure 3: the MSDP of the Pic du Midi Turret Dome (11 channels, NaD1 line, solar granulation).*

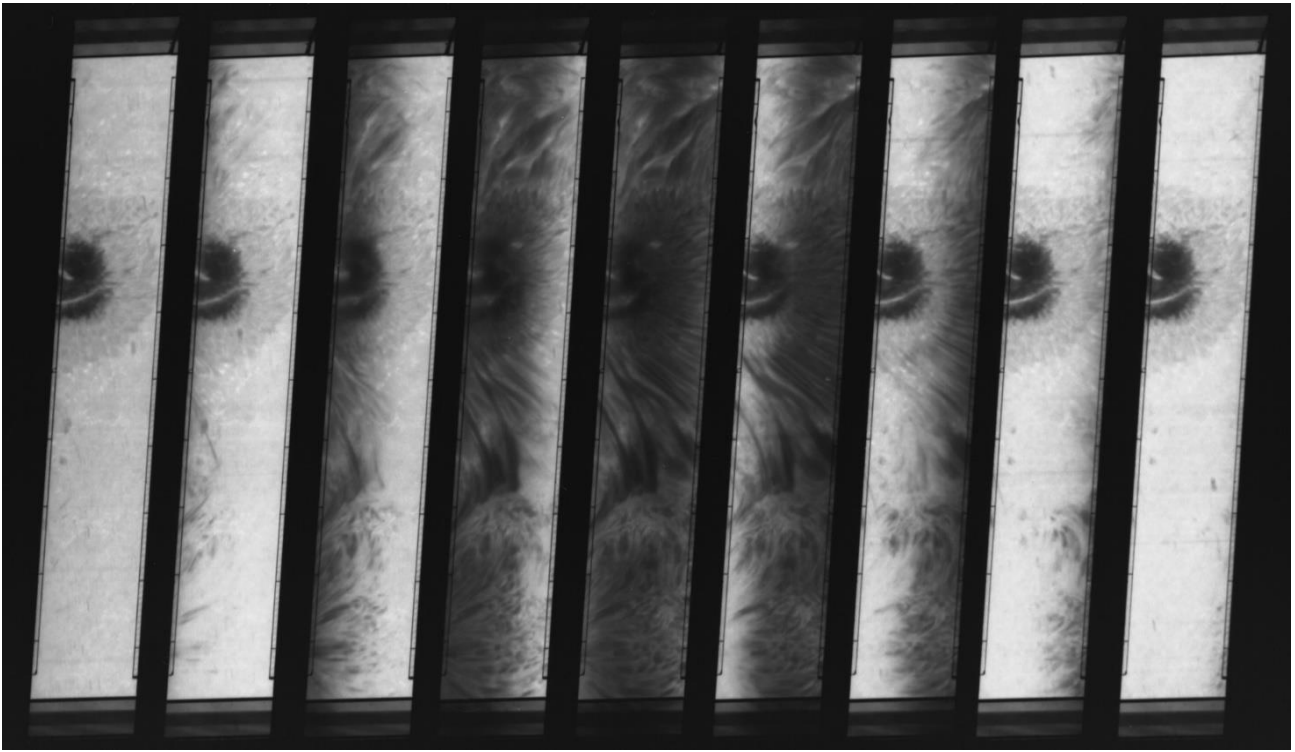

*Figure 4*: *the MSDP of the Vacuum Tower Telescope (9 channels, Hα line, sunspot and active region).*

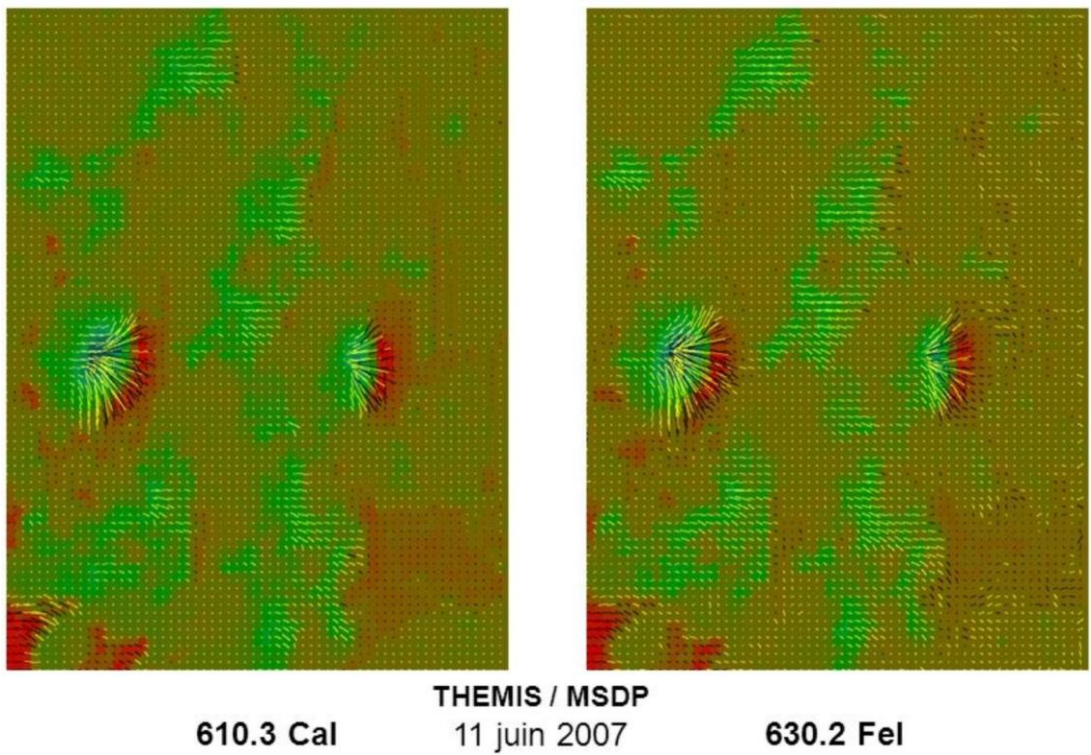

*Figure 5*: *some results of the MSDP of the THEMIS telescope in polarimetric mode, in CaI 6103 and FeI 6302 lines (Mein et al, 2009). The line of sight magnetic field is shown in red and green colours. The transverse magnetic field is indicated by the yellow vectors. For this observation, the polarimeter worked in full Stokes mode (I, Q, U, V) with the 16-channel slicer (80 mÅ resolution).*

Il must be noticed, in case of observations of a stabilized FOV of the Sun (with efficient adaptive optics) that the spectral resolution can be doubled with two successive steps (rotations) of the grating (of half a channel); this method would provide, for instance, 32 interlaced channels (40 mÅ resolution) from two sequential observations with the 16-channel (80 mÅ native) slicer of the THEMIS telescope.

The spectral resolution of the MSDP (Δx/d) depends of the slicer step (Δx in mm) and the spectrograph dispersion (d in mm/Å). It is displayed in Figure 6 for THEMIS (0.4 mm step) and more recent micromirrors based slicers (Meudon, EST project). Two slicers (Meudon, 0.3 mm step ; EST, 0.18 mm step) are convenient for thin lines of the photosphere, providing spectral resolutions (Δx/d) in the range 25 to 50 mÅ from the blue to the near IR. Others (THEMIS, 0.4 mm step ; EST, 0.27 mm step) are designed for larger chromospheric lines, with resolutions in the range 50 to 100 mÅ.

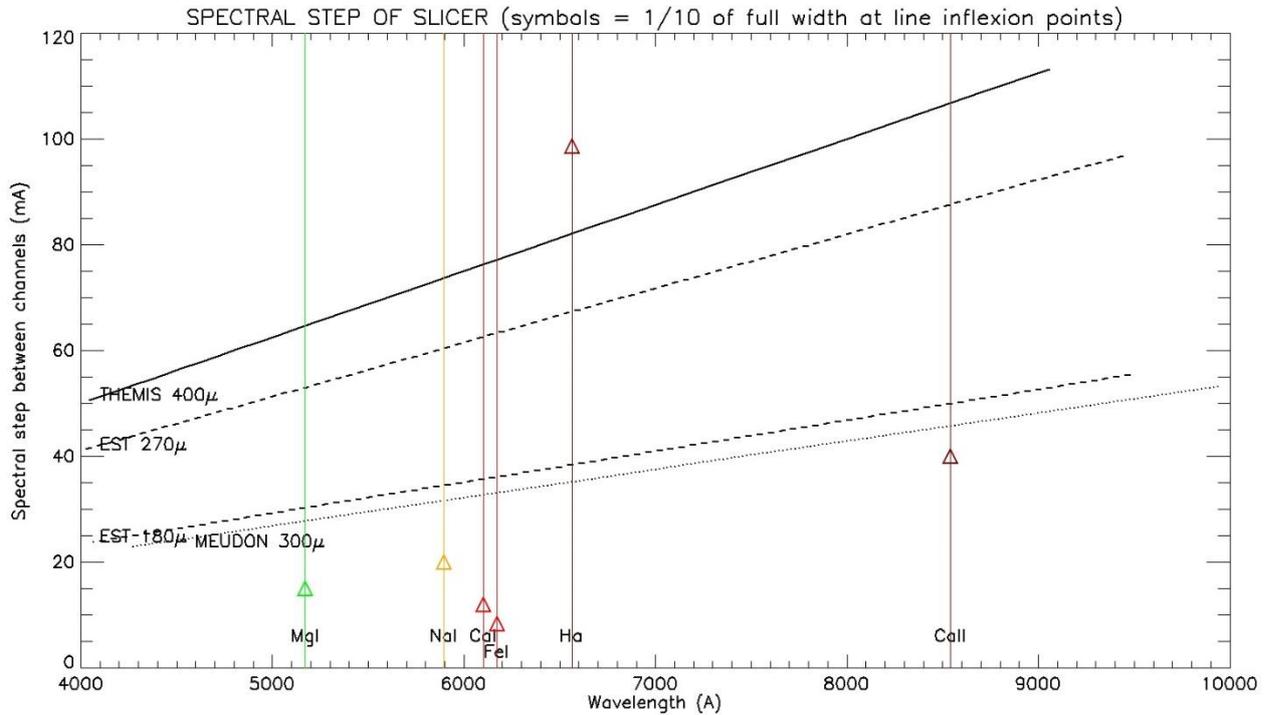

*Figure 6: Spectral resolution (mÅ) of slicers for THEMIS, Meudon and possible MSDP EST, as a function of wavelength. Some classical lines (MgI 5173, NaD1 5896, CaI 6103, FeI 6173, H 6563 and CaII 8542, with respective Full Width at Inflexion Points (FWIP) of 0.15, 0.20, 0.12, 0.08, 1.0 and 0.4 Å) are reported. The triangles indicate the FWIP, divided by 10, of the lines. Thin lines (such as CaI 6103 and FeI 6173) require a high resolution slicer such as EST (0.18 mm) or Meudon (0.30 mm). Broader lines (NaD1, CaII 8542) can be observed with moderate resolution slicers (THEMIS 0.40 mm, EST 0.27 mm).*

However, the spectral resolution is not the only criterion to consider. The spectral bandwidh available around the line core for each solar pixel, whatever its location is in the x-direction of the FOV, is extremely important, especially in the case of large Dopplershifts. The line centering over the channels depends on the abscissa (x) in the FOV (as shown by the graphs of Figure 2 or the MSDP images of Figures 3 and 4). High resolution slicers (Δx/d small), such as the future EST slicer (0.18 mm step), must therefore have a large number of channels (N = 56) to cover the line everywhere in the x-direction. This is explained by Figures 7 and 8, in which is reported the bandwidth available around the line cores over the full FOV (0 < x < $x_m$) of the MSDP window (Figure 7), or the half FOV ( ¼ $x_m$ < x < ¾ $x_m$, Figure 8). Six lines are shown (MgI 5173, NaD1 5896, CaI 6103, FeI 6173, H 6563 and CaII 8542) and, for each, the bars cover a spectral range from 0 to 2 FWIP (a reasonable value to study the line profiles) ; this spectral coverage is chosen to observe properly the line core, the inflexion points, and the close wings. For instance, the Hα line has 1.0 Å FWIP. The corresponding bar covers the interval 0.0-2.0 Å; 1.0 Å bandwidth is the minimum to record the line core, but 2.0 Å bandwidth is necessary to observe inflexion points and close wings, in case of Dopplershifts or in order to probe the chromosphere at several altitudes. Hence, in terms of wavelength coverage only, THEMIS 0.4 mm and EST 0.18 mm are perfect over the full FOV, except for Hα which requires more wavelength coverage. The bandwidth of Meudon is good for thin CaI 6103 and FeI 6173 lines only. If one considers the half FOV, the wavelength coverage becomes larger (factor 1.5 to 2) and is convenient with most slicers; however, working with the half FOV will decrease 2 times the temporal resolution in the case of large x-direction scans of the solar surface, because the spatial step of the scan will be reduced.

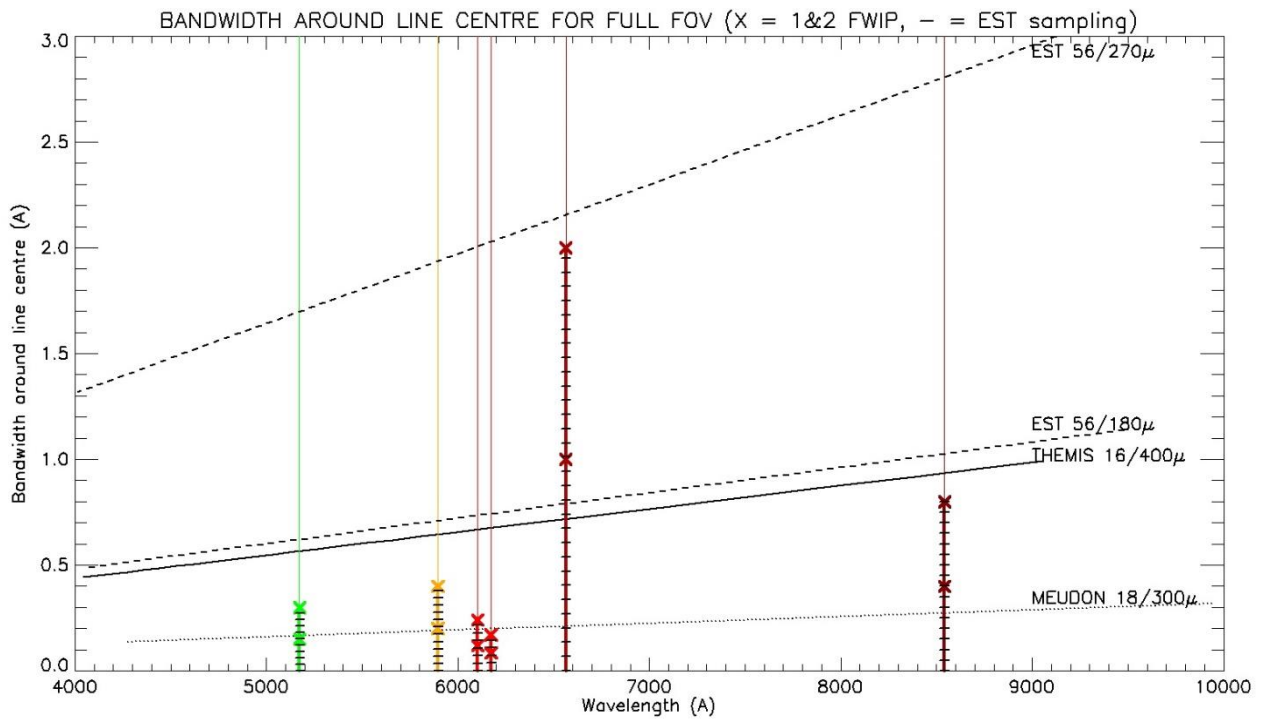

*Figure 7* : Bandwidth (Å) available around the line cores as a function of wavelength, at any point of the **full** FOV (in x-direction). Coloured bars represent MgI 5173, NaD1 5896, CaI 6103, FeI 6173, Hα 6563 and CaII 8542 lines, their length is equal to 2 FWIP. The black dashes on the bars show the MSDP **EST** sampling of the 0.18 mm slicer.

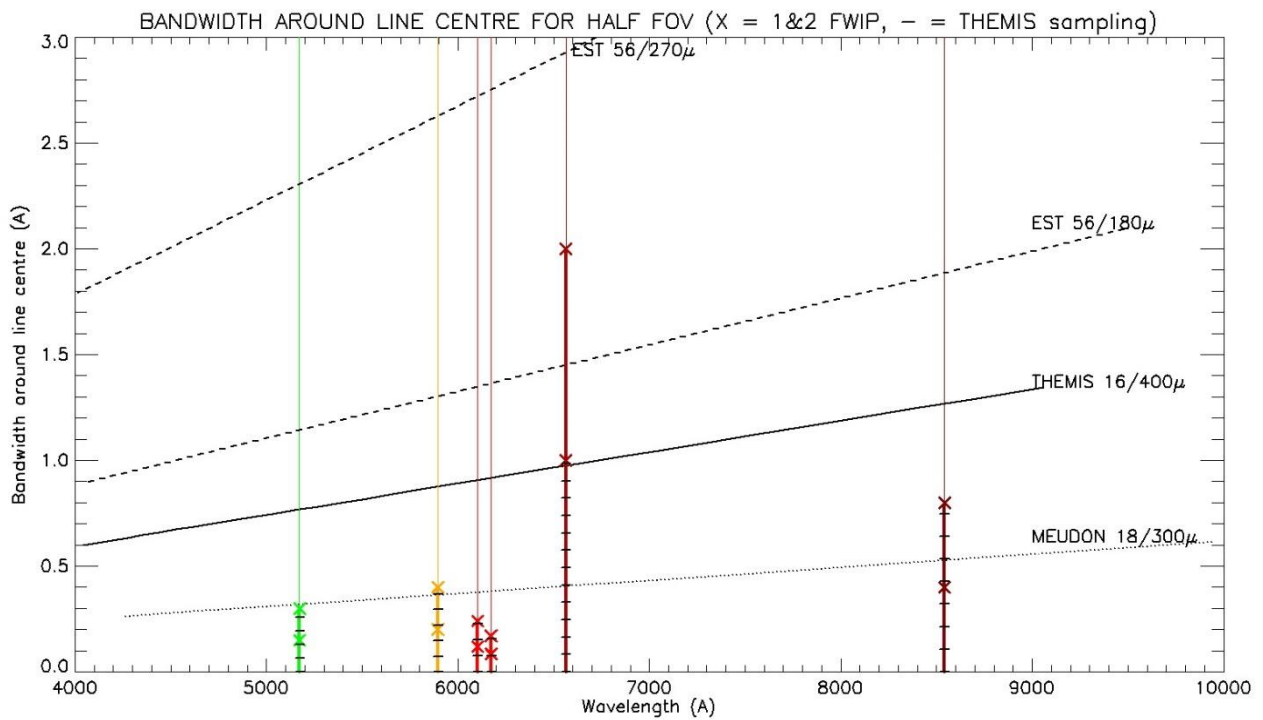

*Figure 8* : Bandwidth (Å) available around the line cores as a function of wavelength, at any point of the **half** FOV (in x-direction). Coloured bars represent MgI 5173, NaD1 5896, CaI 6103, FeI 6173, Hα 6563 and CaII 8542 lines, their length is equal to 2 FWIP. The black dashes on the bars show the MSDP **THEMIS** sampling of the 0.4 mm slicer.

## 2 – The MSDP proposal for EST: the optical design

The integration of the MSDP system into the spectrographs was discussed by Sayède (2023). It uses the 32" x 30" FOV of the telescope at F/50, but the beam is adapted in terms of focal aperture at F/40 (160 m focal length, solar diameter 1488 mm) and re-arranged to fit with the rectangular MSDP entrance window, as shown by Figure 9 ; the initial square FOV is transformed into a long rectangular FOV of 8" x 120" (6.2 x 93.0 mm²) by four image slicers and injected into the spectrographs (Figure 10).

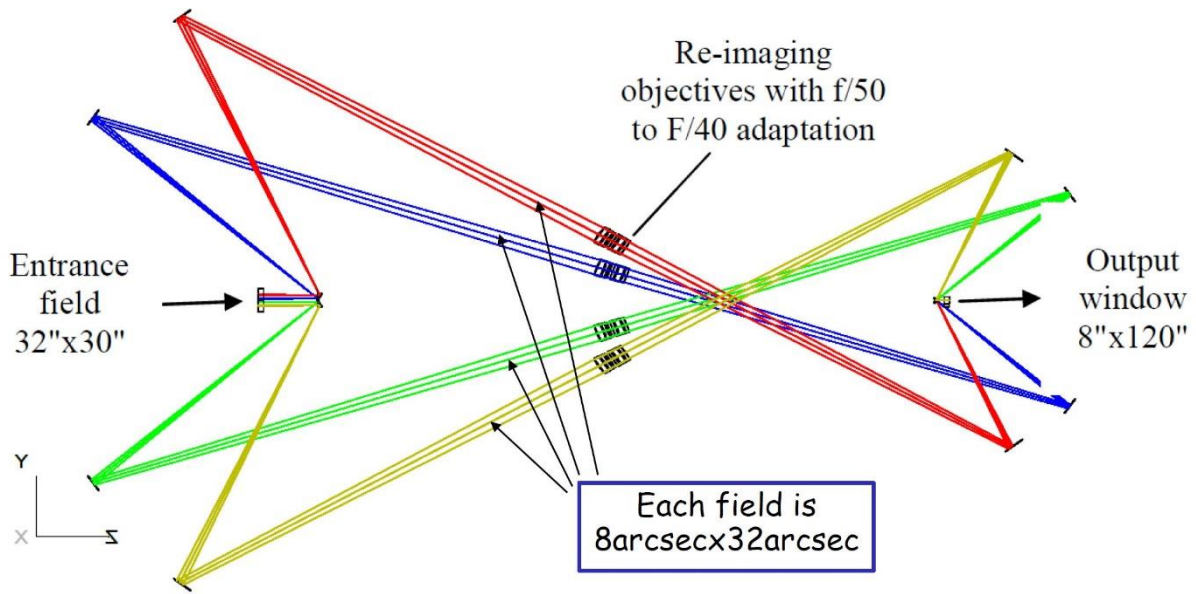

*Figure 9* : The re-imaging system transforms the 32" x 30" entrance FOV into the 8" x 120" output window adapted to the rectangular FOV of the MSDP device. The polarimeter, if any, must be located at the entrance FOV at F/50 in order to avoid instrumental contamination by the re-imaging system (after Sayède, 2023).

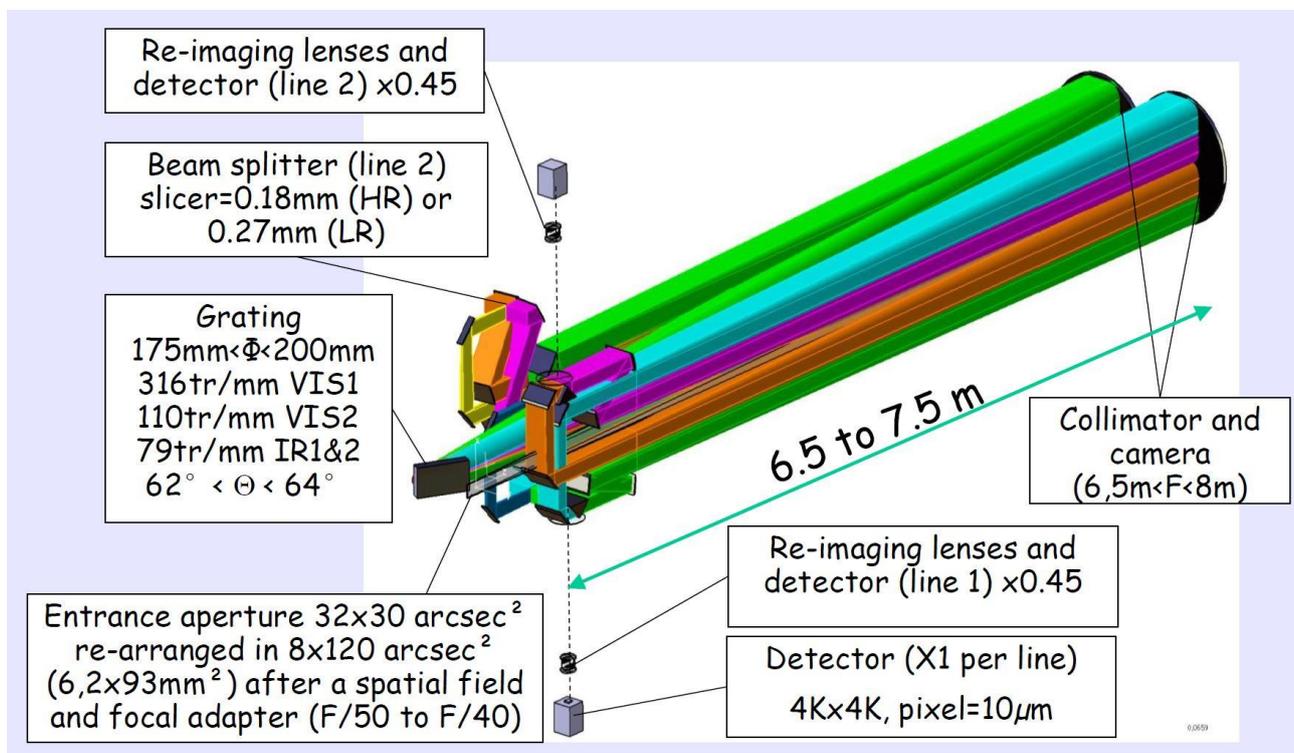

*Figure 10* : The two MSDP slicers (0.18 and 0.27 mm step, for thin and broad lines) are incorporated into the EST spectrographs and use either 4K x 4K detectors (pixel size 0.06" = 10 μ) or 8K x 8K detectors (pixel size 0.03" = 5 μ). Several gratings may be used for visible and IR lines (after Sayède, 2023).

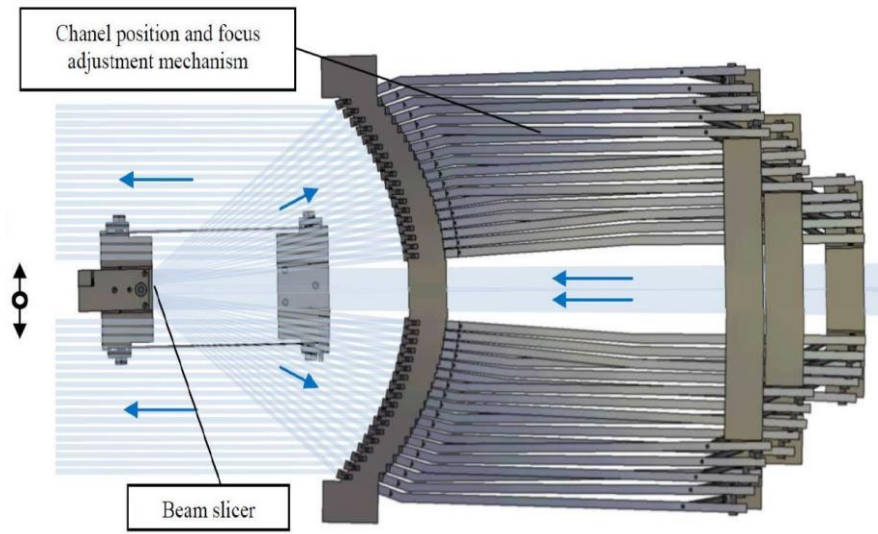

*Figure 11: The MSDP slicers, for thin and broad lines, are made of 56 splitting micro-mirrors (0.18 or 0.27 mm step) and 56 associated adjustable shifting-mirrors (6.7 mm shift) providing at the focus of the spectrograph 56-channel spectra images of about 188 x 188 mm² (after Sayède, 2023).*

| Spectrograph | VIS1 | VIS2 | IR |
|---|---|---|---|
| Wavelength range (nm) | 390-560 | 560-1100 | 700-1600 |
| focal length (m) | 7.5 | 7.0 | 8.0 |
| grating (grooves/mm) | 316 | 110 | 79 |
| blaze angle (°) | 63 | 64 | 62 |
| nλ (A) | 56400 | 163400 | 223500 |
| Examples of order/dispersion (mm/A) for three spectral lines | | | |
| MgI 5173 | 11/5.74 | | |
| FeI 6173 | | 26/4.57 | |
| CaII 8542 | | 19/3.34 | 26/3.50 |
| Full Width at Inflexion Points (FWIP, mA) | | | |
| MgI 5173 | 82 | | |
| FeI 6173 | | 84 | |
| CaII 8542 | | | 400 |
| Wavelength resolution (mA) with 0.18 mm slicer step | | | |
| MgI 5173 | 31 | | |
| FeI 6173 | | 39 | |
| CaII 8542 | | 54 | 51 |
| Wavelength resolution (mA) with 0.27 mm slicer step | | | |
| MgI 5173 | 47 | | |
| FeI 6173 | | 59 | |
| CaII 8542 | | 81 | 77 |
| R with 0.18 mm slicer step | 167000 | | |
| R with 0.27 mm slicer step | 110000 | | |
| Optional slicer : R with 0.10 mm step | 300000 | | |

*Table 2: optical capabilities of the MSDP proposal for EST*

The MSDP proposal for EST consists in two 56-channel slicers (Figure 11), 8" x 120" FOV, for thin and broad lines, available from the visible to the IR. Optical capabilities are summarized by Table 2. Each slicer is made of 56 splitting micro-mirrors and 56 associated shifting and adjustable large mirrors.

## 3 – The MSDP proposal for EST: a simulation of observations

The 32" x 30" FOV is rearranged by the system of Figure 9 into a rectangular FOV of 8" x 120".

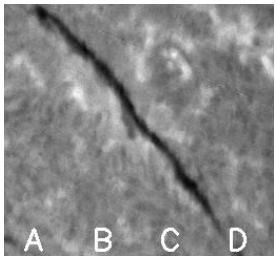

A B C D (32'' x 30'') → A (8'' x 120'')
Square FOV         B   long rectangular FOV
                   C   to deal with MSDP entrance window
                   D

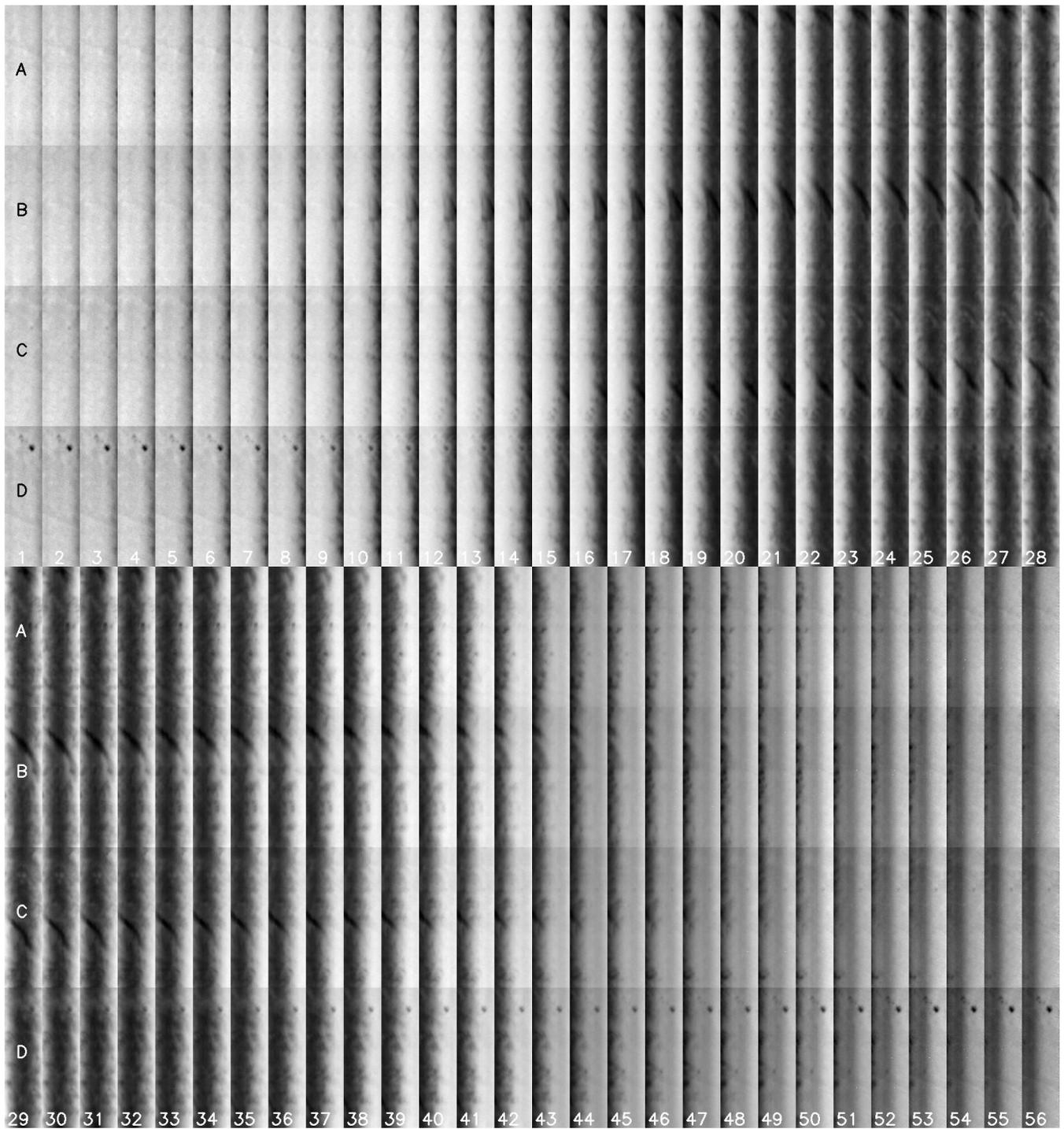

*Figure 12*: *56-channel MSDP spectra image of Hα (0.27 mm slicer), the wavelength step between channels is 61 mÅ. The width is 8". The height is 4 x 30" = 120".*

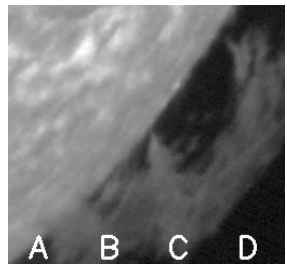

A B C D → A
 B
 C
 D

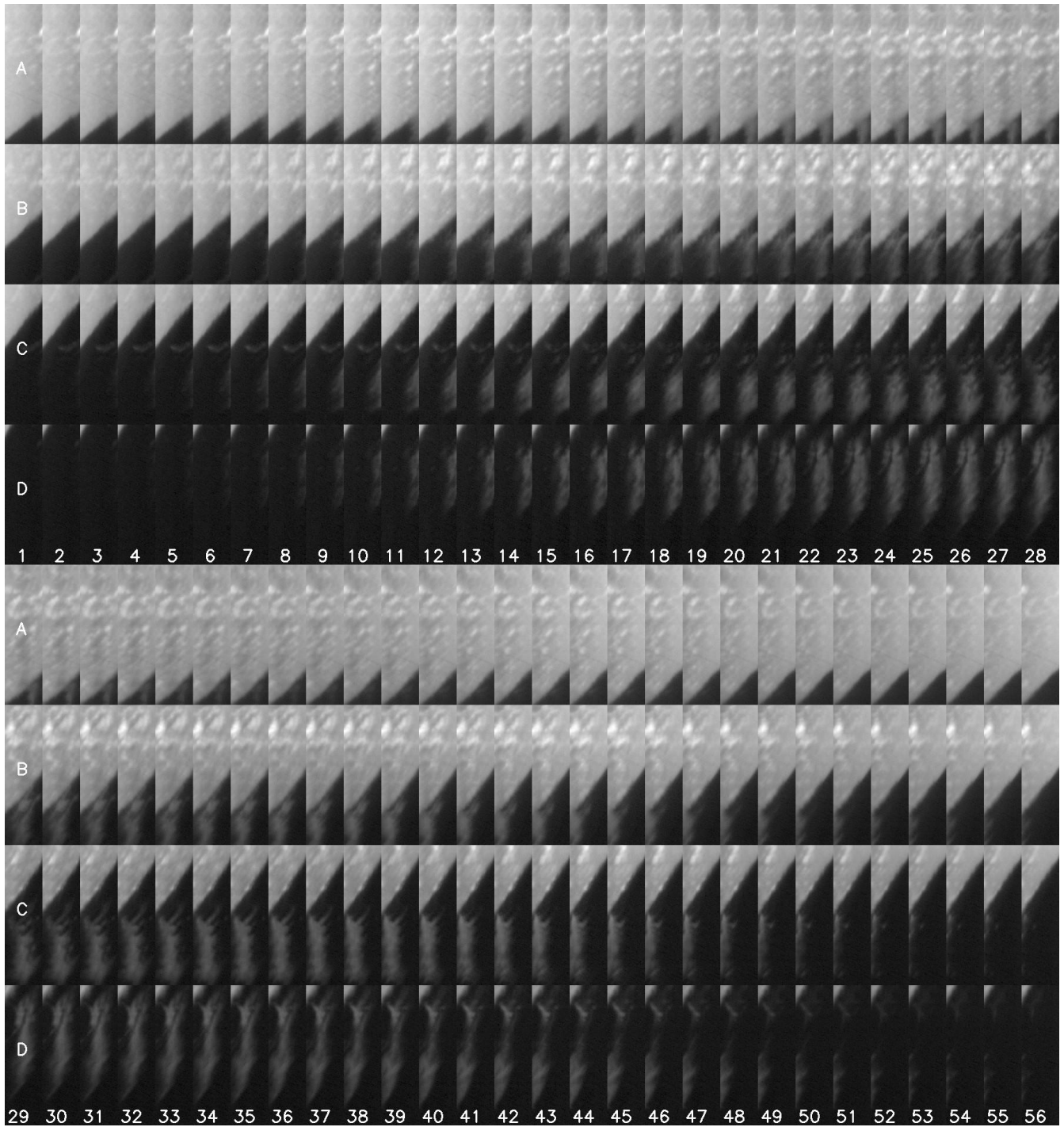

*Figure 13*: 56-channel MSDP spectra image of CaII K (0.18 mm slicer), the wavelength step between channels is 25 mÅ. The width is 8". The height is 4 x 30" = 120".

The 56-channel spectra images are re-arranged in a square format of about 18 x 18 cm² displaying one group of 28 channels top, and the second group of 28 channels bottom (Figures 12 and 13). Figure 14 shows line profiles got at 3 locations in the x-direction of the FOV, x = 0 (left), $x_m/2$ (centre), $x_m$ (right).

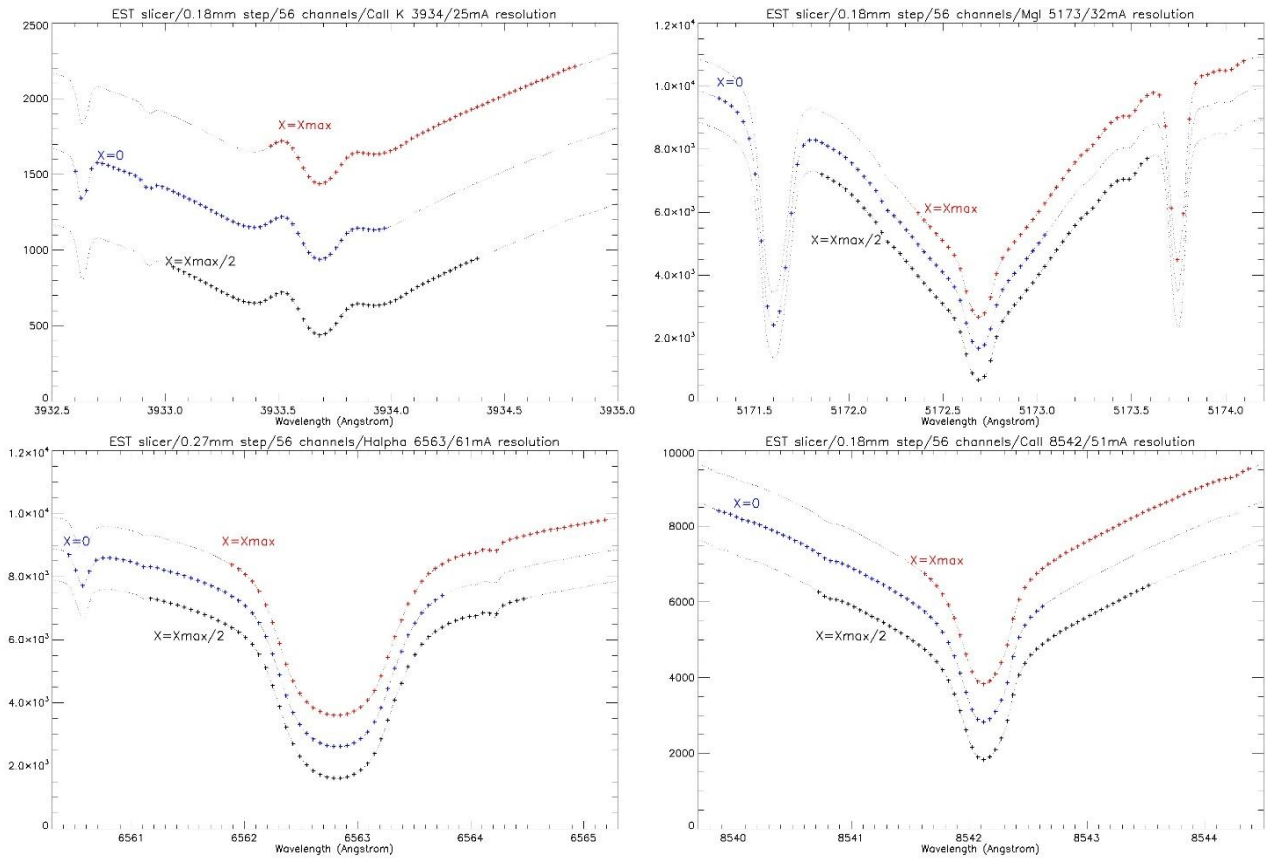

*Figure 14* : CaII K, MgI 5173, Hα and CaII 8542 line profiles, sampled by the 56-channel slicers, for 3 locations in the x-direction of the FOV (left, centre, right). The spectral resolution for CaII K, MgI 5173 and CaII 8542 is, respectively, 25, 32, 51 mÅ with the 0.18 mm slicer. For Hα, it is 61 mÅ with the 0.27 mm slicer.

**4 – The MSDP proposal for EST with polarimetry**

The MSDP is compatible with most polarimetric methods. However, the polarization analysis must be achieved at the F/50 focus (32" x 30" square FOV), because the beam adaptation from F/50 to F/40 and the image re-alignment (32" x 30" → 8" x 120") uses flat mirrors which may introduce instrumental polarization. Among the numerous polarimetric methods, we can emphasize at least 2 methods:

1) High speed **modulation** (100 Hz) of a **single beam** with adaptive optics allowing fast and sequential measurements of I+Q, I-Q, I+U, I-U, I+V, I-V of the stabilized 32" x 30" FOV. In that case, **the MSDP can be used as it is**, there is no need of extra hardware. The limitation could be the exposure time (< 10 ms with 100 Hz modulation) which could reduce the photon flux. This method was used by Roudier et al (2006) at Pic du Midi, with moderate success, because the cadence was slow (5 Hz) and there was no image stabilization (they used post processing image selection and destretching).

2) If one needs strictly simultaneous measurements of I+S, I-S (where S = Q, U, V), a **dual beam** is required. The 32" x 30" FOV can be split in two parts with a grid located at the F/50 focus followed by a calcite polarizing beam splitter (Semel, 1980, Figure 15). The shift can be either in the X-direction (Meudon Solar Tower) or in the Y-direction (THEMIS telescope), as shown by Figure 16. Both methods present advantages and disadvantages. The X-direction split simply reduces the MSDP FOV from 8" x 120" to 4" x 120", but I+S and I-S profiles are not sampled at the same wavelengths. Interpolations are required before combining them. Alternatively, the Y-direction split divides the 8" x 120" FOV in tens of 8" x 4" elementary zones ; but in that case, I+S and I-S profiles are sampled at the same wavelengths. At the F/50 focus, 4" = 3.87 mm, a 35 mm thickness (or 2 * 17.5 mm) calcite beam splitter is required with X, Y dimensions of 32" x 30" = 31 x 29 mm². This volume (31 x 29 x 35 mm$^3$) is quite reasonable. At the MSDP F/40 focus, 4" = 3.10 mm, a 28 mm thickness polarizing beam splitter is convenient, but the X, Y dimensions are now very different, 8" x 120" = 6.2 x 93.0 mm². The corresponding volume of calcite (6.2 x 93 x 28 mm³) may be more difficult to produce. We now discuss the MSDP polarimetry using a polarizing beam splitter acting either in X or Y directions.

The splitting principle is summarized by Figure 15. A grid is located in the image plane at the F/50 focus; the step of the grid is 8" (4" open, 4" closed), so that the grid transmits the half FOV. In order to observe the full FOV, two successive observations are needed for two positions of the solar image, shifted of 4". The grid is followed by two birefringent glasses arranged to equalize the optical paths ; it transmits two simultaneous states of polarization (dual beam) of the same FOV element (for instance, 1 and 1' are cospatial and provide I+S and I-S, S = Q, U, V).

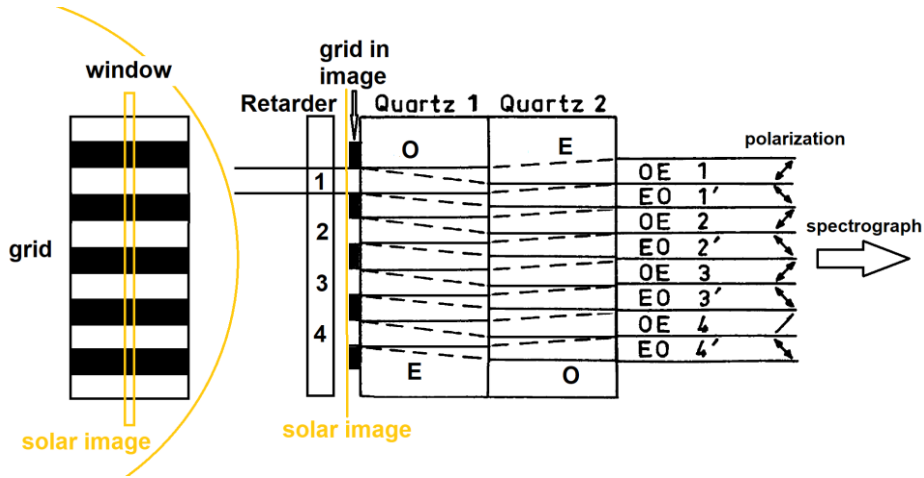

*Figure 15* : Polarimetry with grid (after Semel, 1980). O and E designate the ordinary and extraordinary rays.

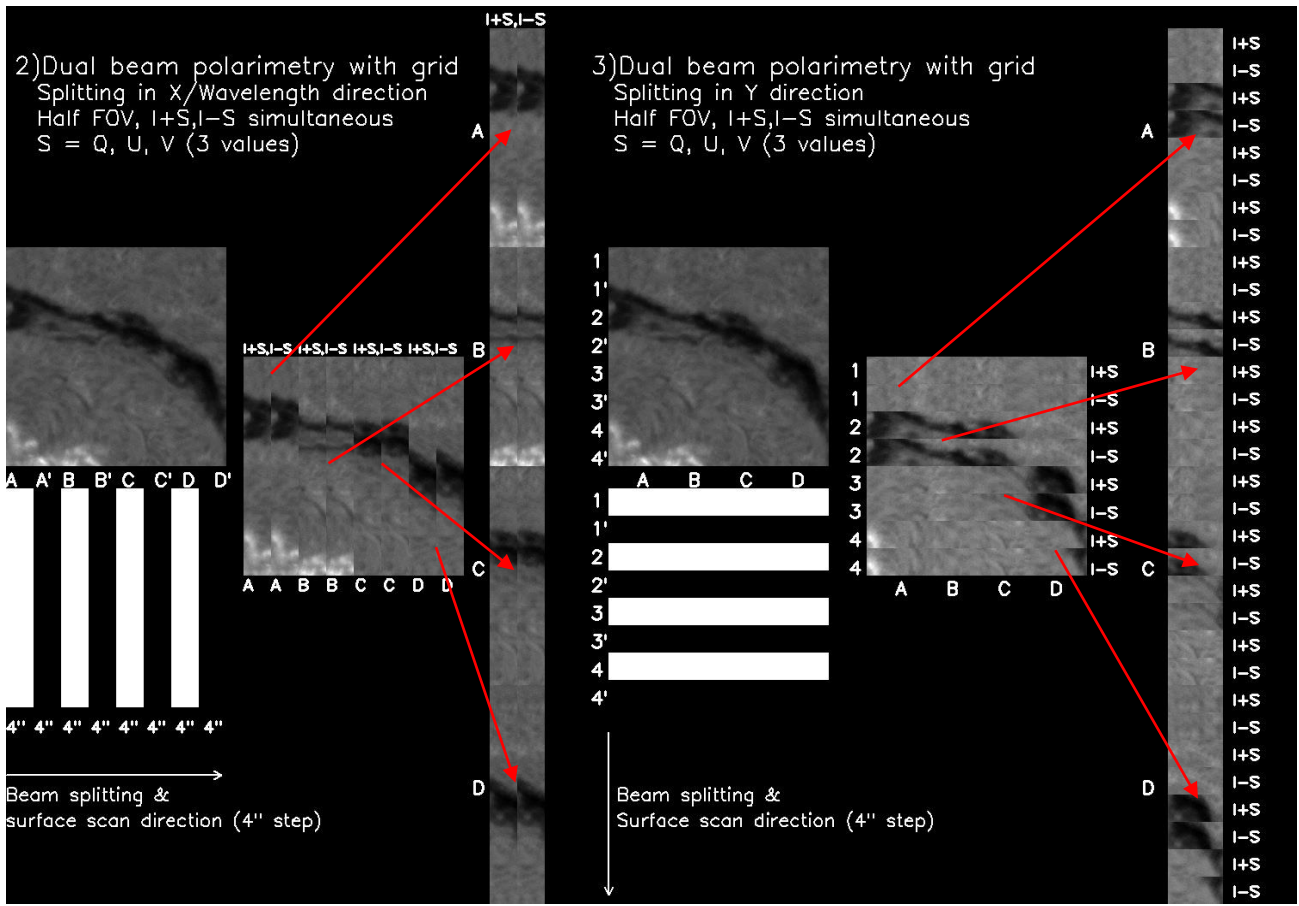

*Figure 16* : MSDP polarimetry with dual beams, beam splitting either in X (left) ot Y (right) directions.

The action of the grid and associated beam-splitter is explained in Figure 16.

When the splitting acts in X-direction:

- The 32" x 30" FOV is made of 8 contiguous regions A, A', B, B', C, C', D, D' of 4" x 30".
- The grid selects regions A, B, C, D of 4" x 30" (half FOV) while regions A', B', C', D' are blocked
- The polarizing beam-splitter provides A, B, C, D in two simultaneous states of polarization I+S, I-S (S = Q, U, V observed in sequence)
- The MSDP transfer optics from F/50 to F/40 rearranges the field such that A, B, C, D are presented in 4 rows, while columns 1-2 are devoted to I+S and I-S

When the splitting acts in Y-direction:

- The 32" x 30" FOV is made of 8 contiguous regions 1, 1', 2, 2', 3, 3', 4, 4' of 32" x 4".
- The grid selects regions 1, 2, 3, 4 of 32" x 4" (half FOV) while regions 1', 2', 3', 4' are blocked
- The polarizing beam-splitter provides 1, 2, 3, 4 in two simultaneous states of polarization I+S, I-S (S = Q, U, V observed in sequence)
- The MSDP transfer optics from F/50 to F/40 rearranges the field such that the small regions (8" x 4") inside A, B, C, D are presented in many rows, alternatively in two states of polarization (I+S and I-S)

Hence, the X-direction splitting produces 4 regions of 4" x 30", each in two states of polarization. On the contrary, the Y-direction splitting gives 16 small area regions of 8" x 4", each again in two states of polarization, but in that case, the initial FOV is much more fragmented.

### 4 – 1 - Y-direction splitting for dual beam

This method was used with great success on the THEMIS telescope (Figure 17).

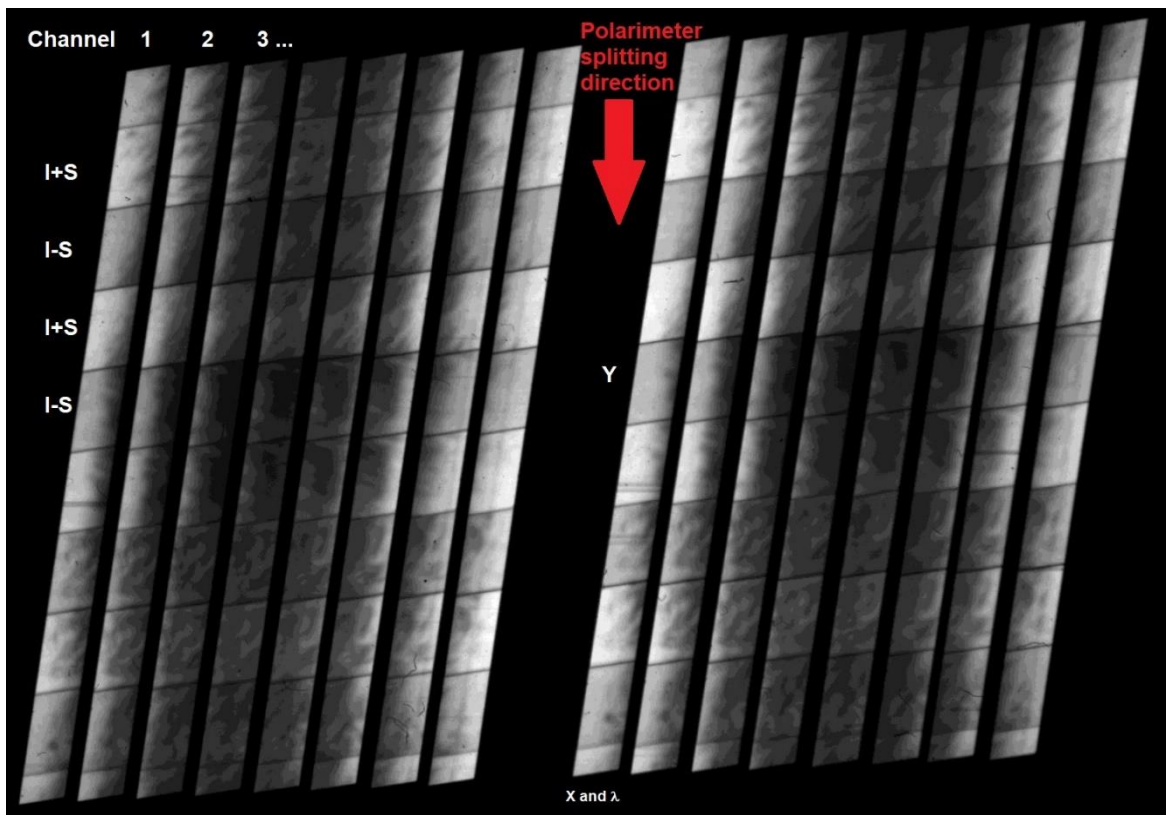

*Figure 17* : *Polarimetric splitting in Y direction with the 16-channel slicer of the THEMIS telescope, Hα line.*

We take the example of MgI 5173 and show below (Figures 18, 19, 20, 21) what could produce the MSDP onboard EST with the 56-channel slicer (0.18 mm step).

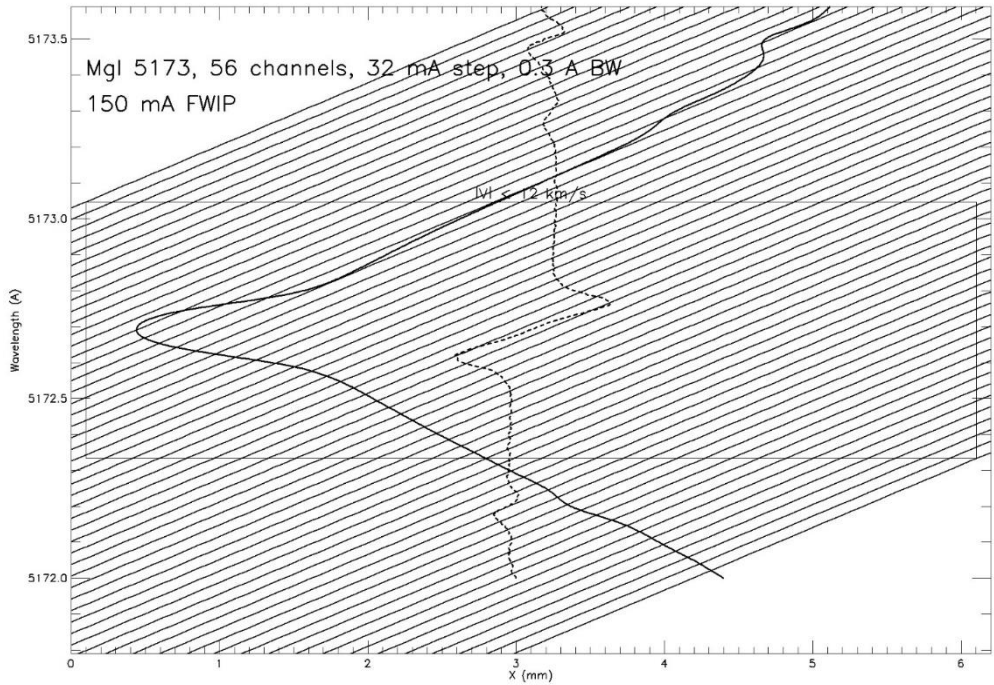

*Figure 18* : the MSDP wavelength transmission $\lambda_n(x)$ for the 56 channels slicer and MgI 5173 line (0.032 Å resolution). MgI 5173 is a thin line (0.15 Å FWIP, for Full Width at Inflexion Points). The line profile at disk centre and its first derivative are displayed. For a given bandpass centred on the line core (0.3 Å), the graph shows that the measure of large Dopplershifts (12 km/s) is possible everywhere in the FOV (6.2 mm = 8").

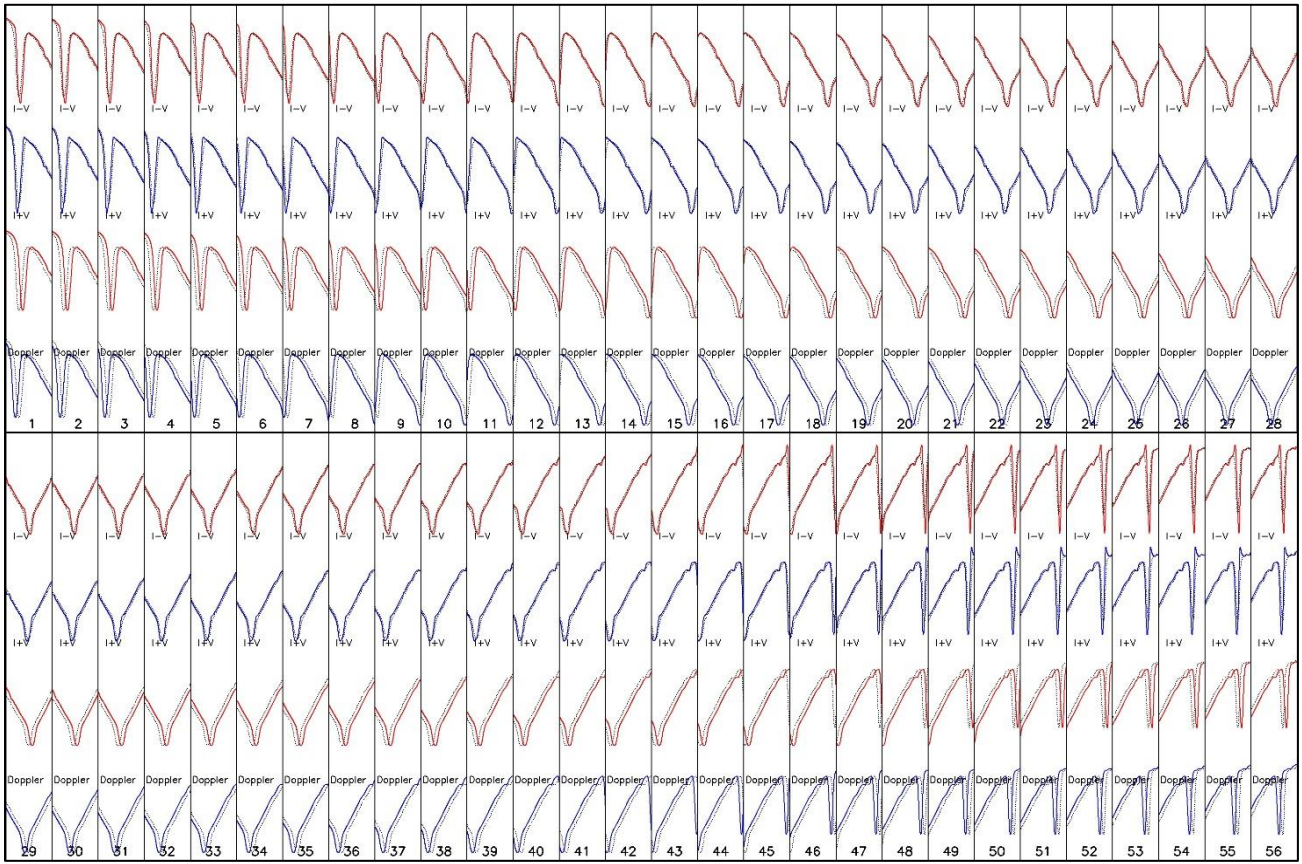

*Figure 19* : Simulation of the 56 channels for MgI 5173 with polarimetric splitting in the Y direction (long dimension of each channel, Semel's grid method). Couples I-Q and I+Q, or I-U and I+U, or I-V and I+V, can be observed in sequence. Here, I-V and I+V are displayed for 1000 G, no velocity. An example of Dopplershifts is also provided for -5 and +5 km/s. The dotted line is the reference profile (no magnetic field, no velocity).

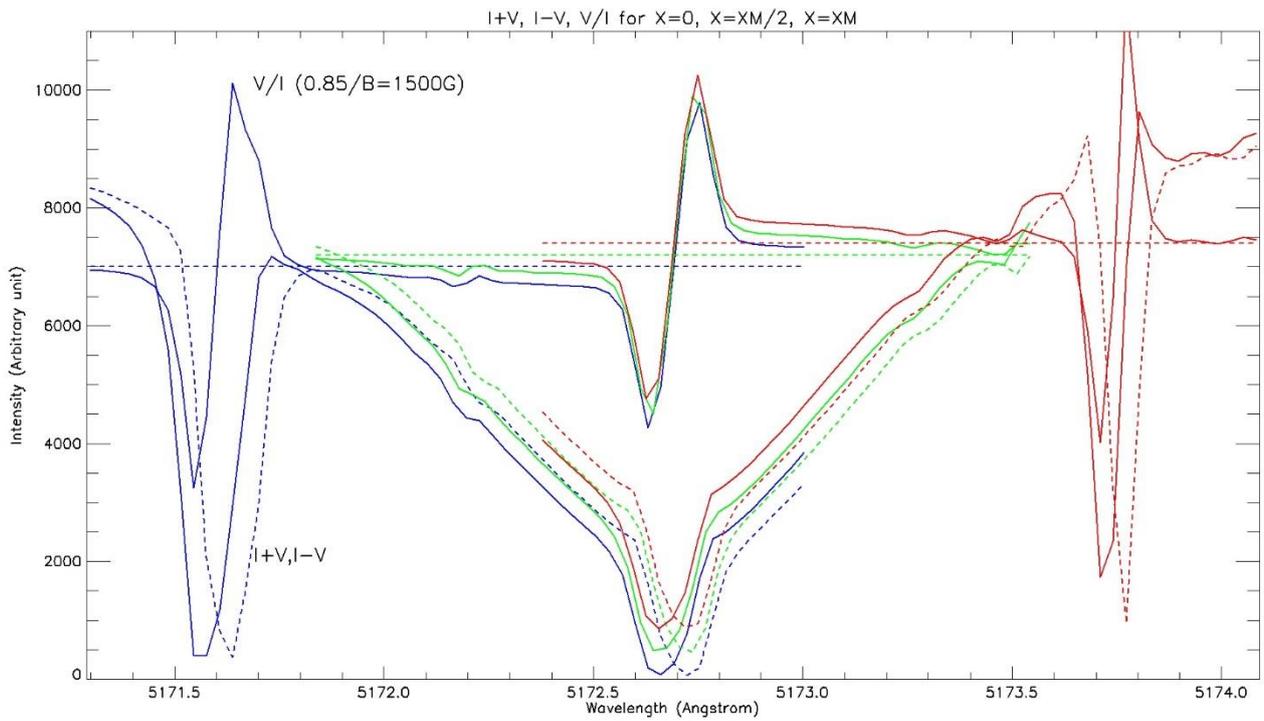

*Figure 20* : Line profiles of I+V and I-V for MgI 5173 with polarimetric splitting in the Y direction (long direction of the channels) that could be got with the 56-channel slicer, together with the polarization rate V/I. Blue, green, red profiles are for 3 locations in the FOV (left, centre, right) corresponding to x = 0, $x_m/2$, $x_m$ ($x_m$ = 6.2 mm corresponds to 8"). Solid line for I+V and dashed line for I-V. The simulation is for B = 1500 G. I+V and I-V profiles are defined by the same sampling points, for a given x-position.

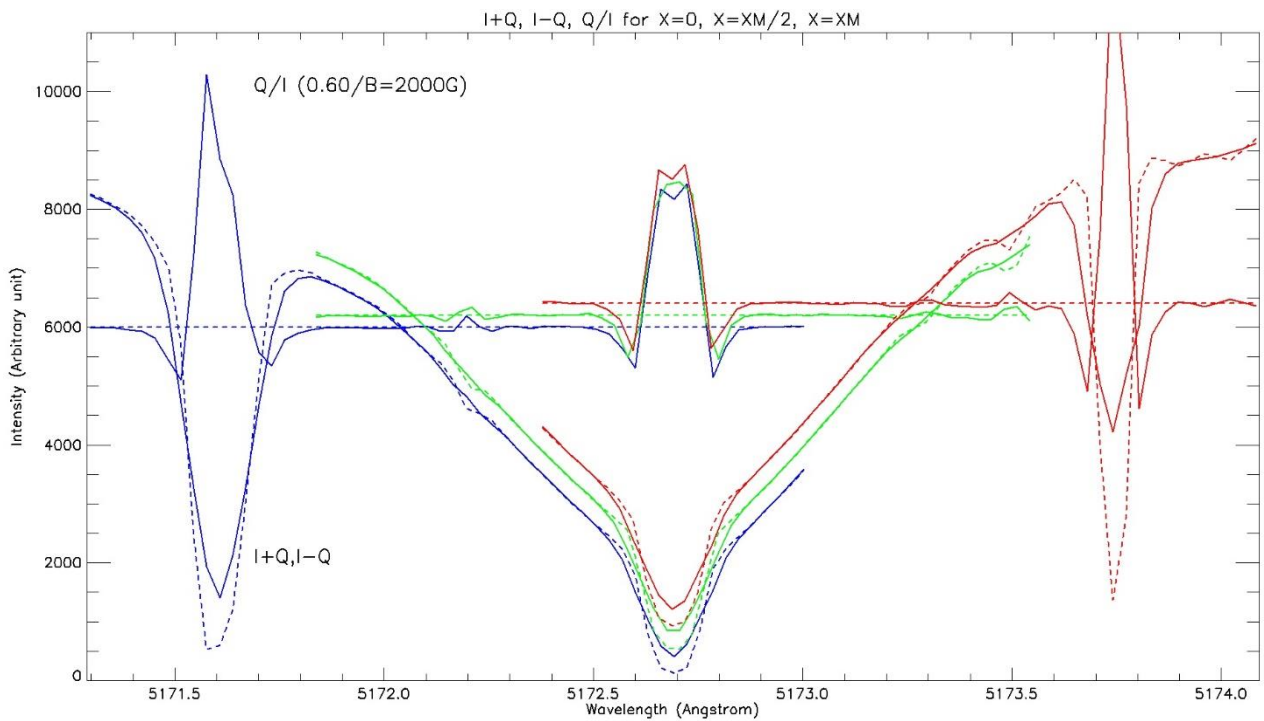

*Figure 21* : Line profiles of I+Q and I-Q for MgI 5173 with polarimetric splitting in the Y direction (long direction of the channels) that could be got with the 56-channel slicer, together with the polarization rate Q/I. Blue, green, red profiles are for 3 locations in the FOV (left, centre, right) corresponding to x = 0, $x_m/2$, $x_m$ ($x_m$ = 6.2 mm corresponds to 8"). Solid line for I+Q and dashed line for I-Q. The simulation is for B = 2000 G. I+Q and I-Q profiles are defined by the same sampling points, for a given x-position.

## 4 – 2 - X-direction splitting for dual beam

This method was used with success at the Meudon Solar Tower (Figure 22).

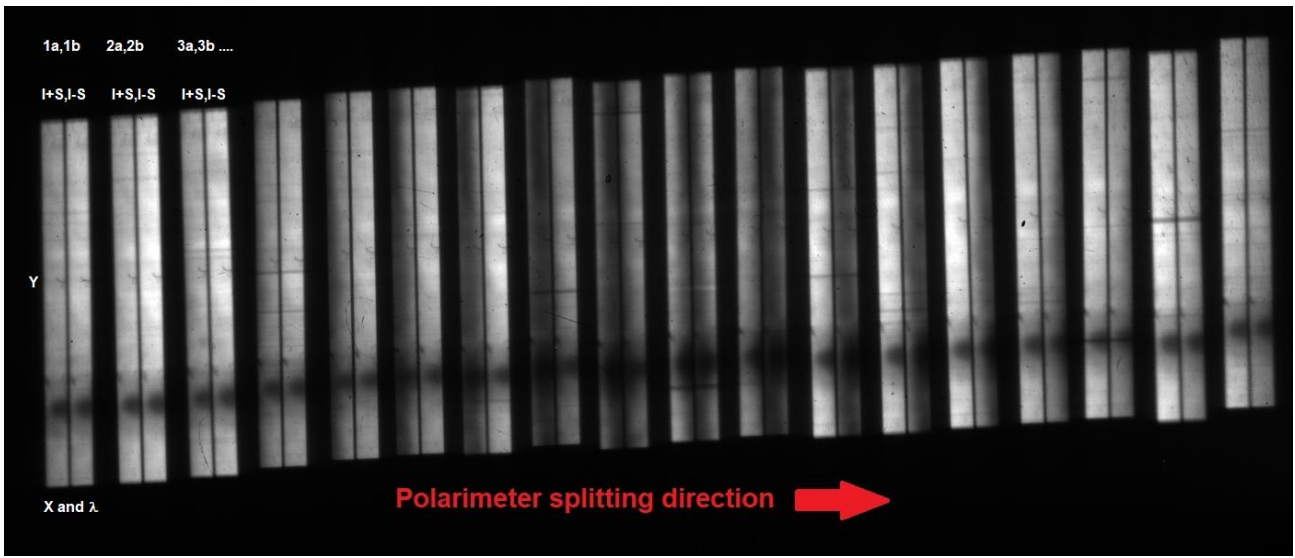

*Figure 22* : Polarimetric splitting in X direction with the 18-channel slicer of the Meudon Solar Tower, MgI 5173 line. Each channel is split into 2 subchannels for I+S, I-S (S = Q, U, V in sequence).

We take the example of MgI 5173 and show below (Figures 23, 24, 25, 26) what could produce the MSDP onboard EST with the 56-channel slicer (0.18 mm step).

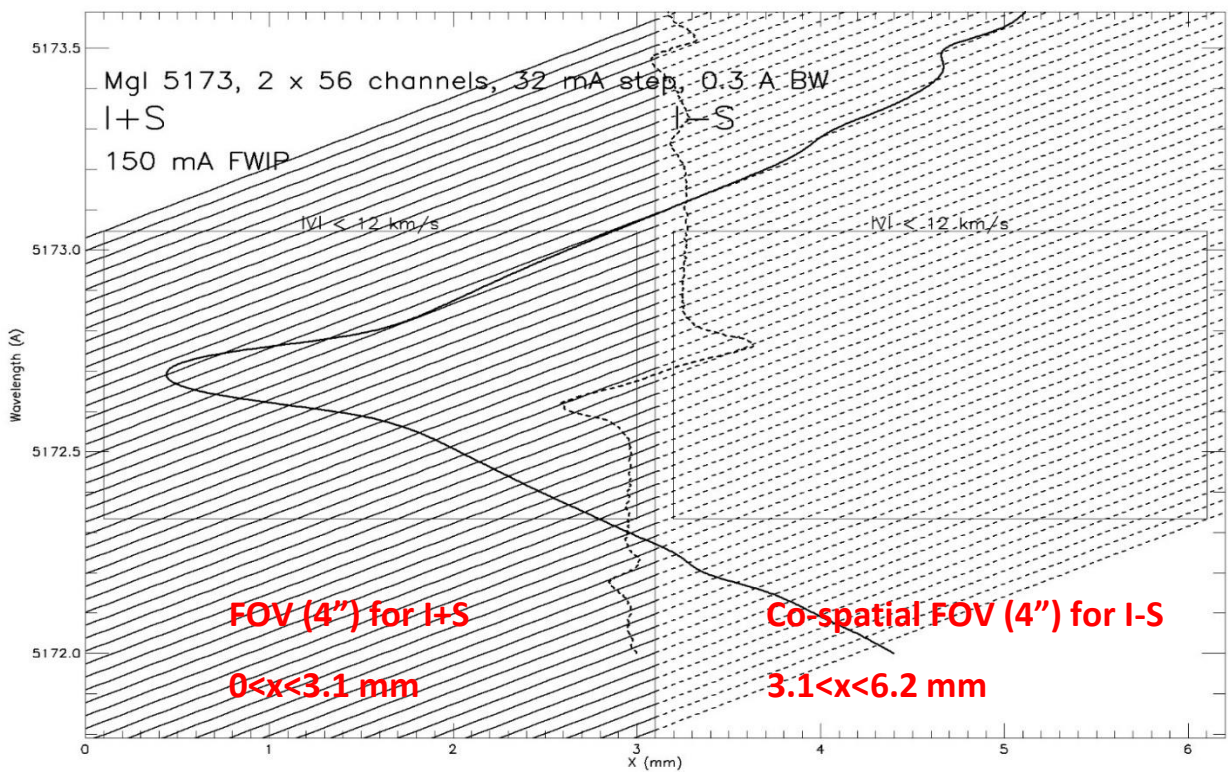

*Figure 23*: the MSDP wavelength transmission $\lambda_n(x)$ for the 56 channels slicer and MgI 5173 line (0.032 Å resolution). MgI 5173 is a thin line (0.15 Å FWIP, for Full Width at Inflexion Points). The line profile at disk centre and its first derivative are displayed. For a given bandpass centred on the line core (0.3 Å), the graph shows that the measure of large Dopplershifts (12 km/s) is possible everywhere in the FOV. Each channel (0<x<6.2 mm) is split into 2 co-spatial subchannels ; profiles I+S(x) and I-S(x+3.1) are co-spatial (0<x<3.1); however, the wavelength sampling points are not the same ; an interpolation is required to combine them.

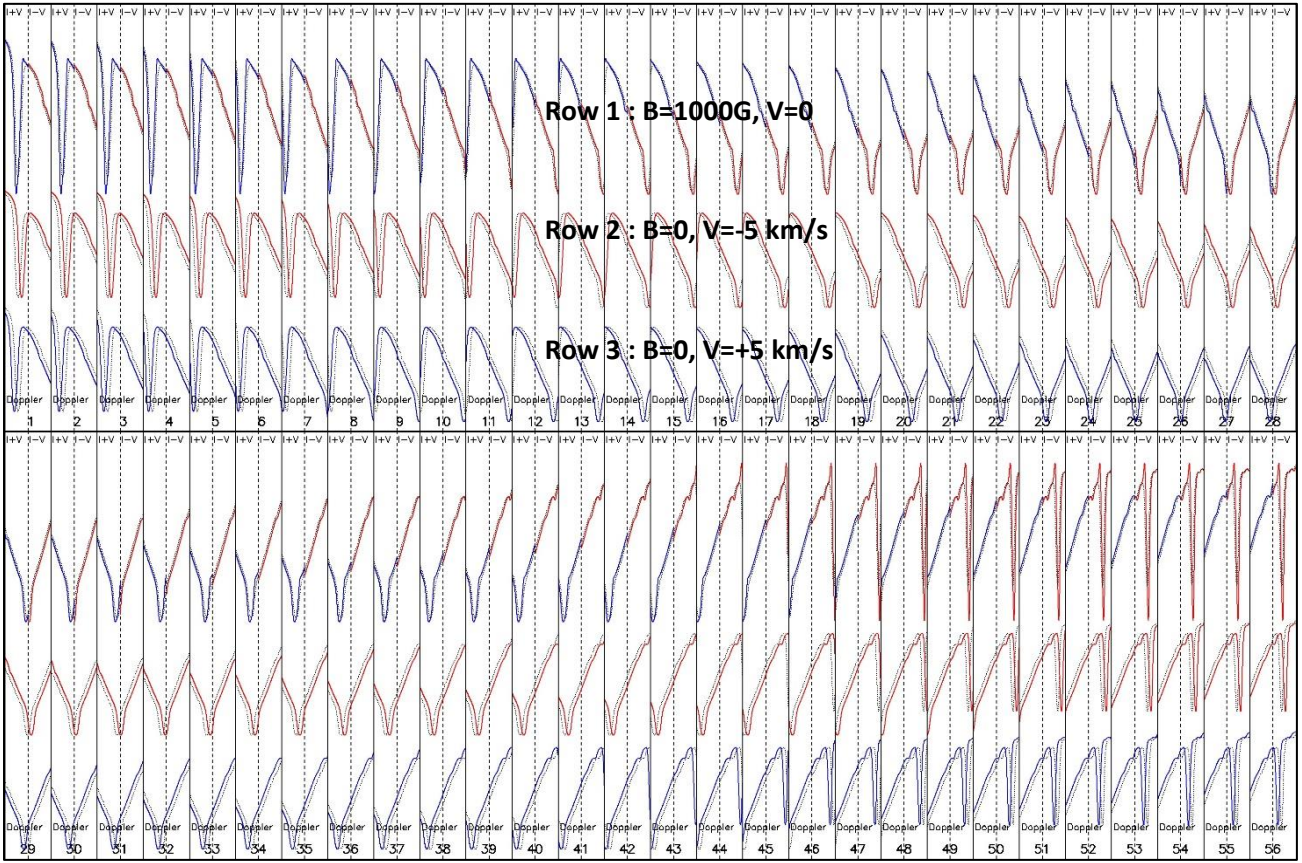

*Figure 24* : *the simulation of the 56 channels for MgI 5173 with polarimetric splitting in the X or wavelength direction (short dimension of each channel), delivering 56 x 2 co-spatial subchannels of half size (4" instead of 8") in X direction. I-S and I+S (S = Q, U, V in sequence) are simultaneous. Here, the V signal is for 1000 G (row 1). An example of Doppershifts is also provided for -5 km/s (row 2) and +5 km/s (row 3). The dotted line is the reference profile (no magnetic field, no velocity).*

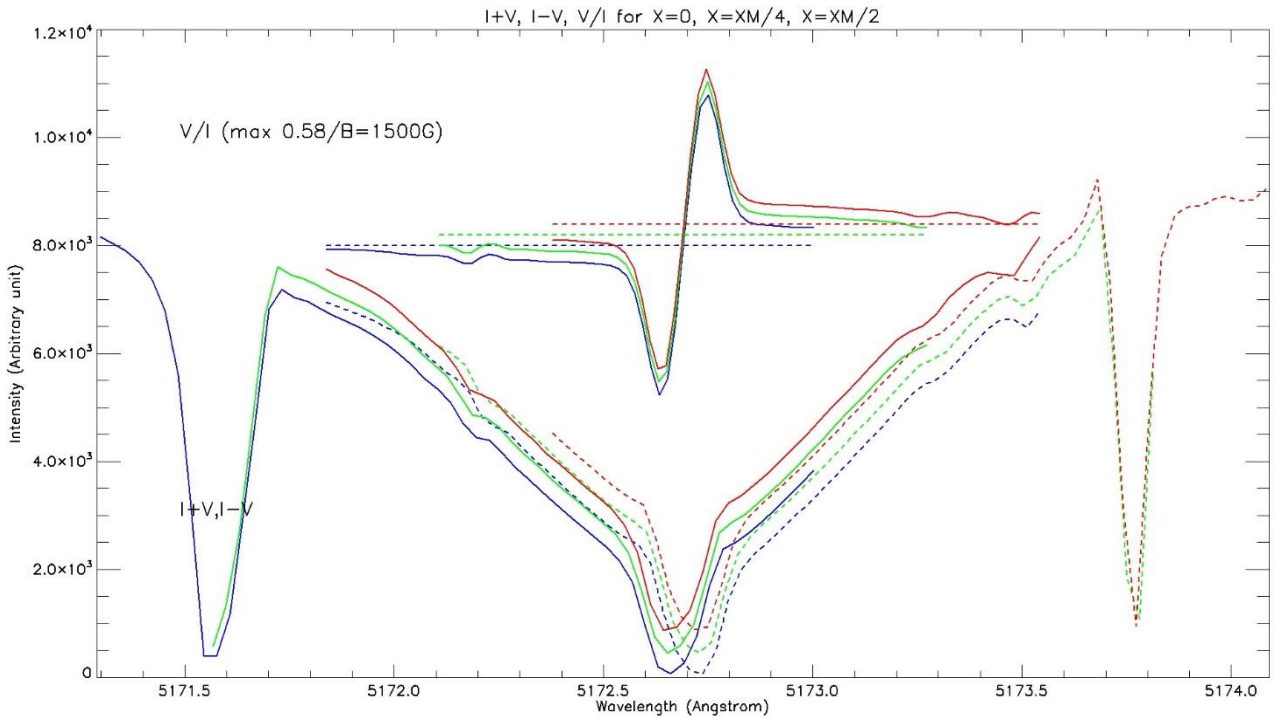

*Figure 25* : *Line profiles of I+V and I-V for MgI 5173 with polarimetric splitting in the X direction (or wavelength direction) that could be got with the 56-channel slicer, together with the polarization rate V/I. Blue, green, red profiles are for 3 locations in the FOV (left, centre, right) corresponding to x = 0, $x_m/4$, $x_m/2$ ($x_m/2$ = 3.1 mm*

corresponds to 4"). Solid line for I+V and dashed line for I-V. The simulation is for B = 1500 G. I+V and I-V profiles are not defined by the same sampling points, for a given x-position, so that the V/I computation, which combines I+V and I-V, requires an interpolation.

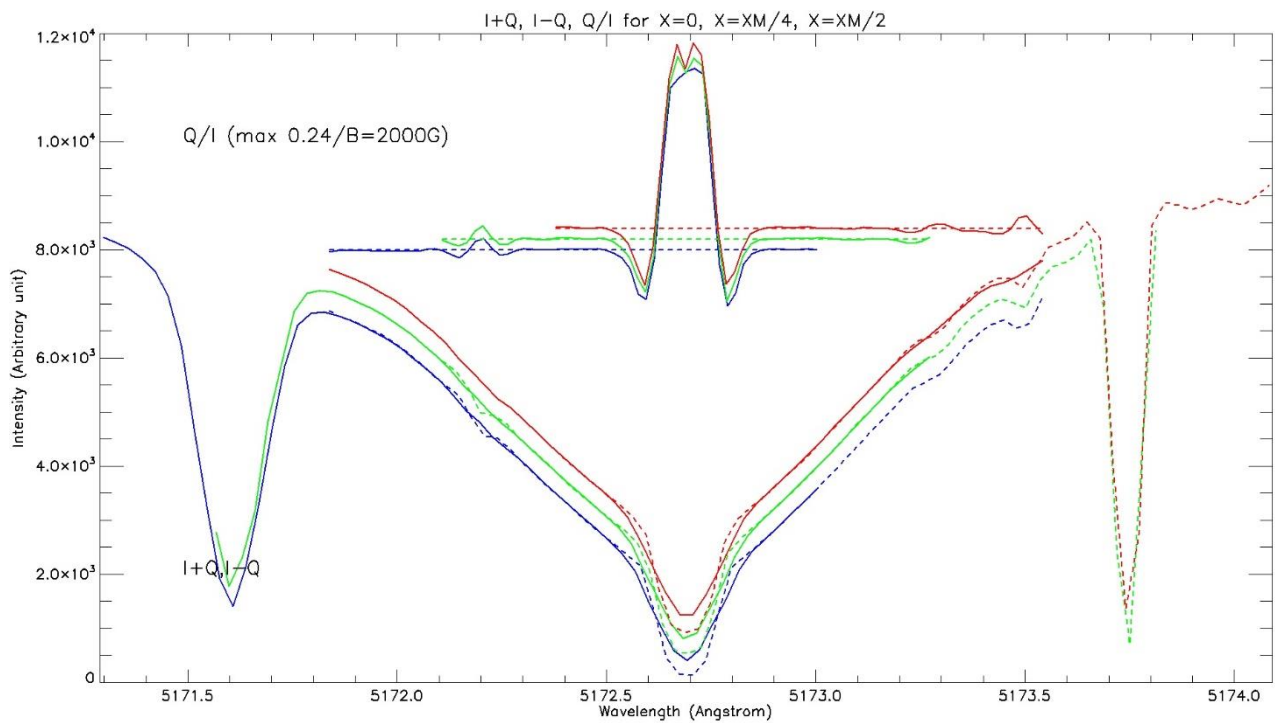

**Figure 26** : Line profiles of I+Q and I-Q for MgI 5173 with polarimetric splitting in the X direction (or wavelength direction) that could be got with the 56-channel slicer, together with the polarization rate Q/I. Blue, green, red profiles are for 3 locations in the FOV (left, centre, right) corresponding to x = 0, $x_m/4$, $x_m/2$ ($x_m/2$ = 3.1 mm corresponds to 4"). Solid line for I+Q and dashed line for I-Q. The simulation is for B = 2000 G. I+Q and I-Q profiles are not defined by the same sampling points, for a given x-position, so that the Q/I computation, which combines I+Q and I-Q, requires an interpolation.

## Conclusion

The MSDP technique, with new technology slicers, providing high spectral resolution and bandwidth, is well adapted for fast imaging spectro-polarimetry of the chromosphere and the photosphere. In comparison to previous MSDP versions, the new slicers improve the spectral resolution (two slicers for EST with R =110000 and R=167000, and the possibility to reach R=300000 on demand), the spectral coverage of lines, the photon flux and the spatial resolution in combination with adaptive optics. The MSDP is a powerful technology that can be incorporated into 7-8 m existing spectrographs. The MSDP is compatible with most polarimetric methods; in particular, high spectral resolution slicers are suitable for full Stokes polarimetry. The high speed modulation technique is very attractive to preserve the FOV (which is often small on large telescopes) ; but the MSDP is also fully compatible with double beam methods (implying FOV reduction and fragmentation) to produce strictly simultaneous Stokes signals.

## References

Malherbe, J.M., Mein, P., Sayède, F., Rudawy, P., Phillips, K., Keenan, F., Rybak, J., 2021, "*The solar line emission dopperometer project*", Experimental Astronomy
https://doi.org/10.1007/s10686-021-09804-x

Mein P., "*Multi-channel subtractive spectrograph and filament observations*", 1977, Sol. Phys., 54, 45.
https://ui.adsabs.harvard.edu/abs/1977SoPh...54...45M

Mein, P., *"Multi-channel Subtractive Double Pass Spectrograph"*, 1980, Japan-France Seminar on Solar Physics, 285
https://ui.adsabs.harvard.edu/abs/1980jfss.conf..285M

Mein P., *"Solar 2D spectroscopy - A new MSDP instrument"*, 1991, Astron. Astrophys., 248, 669.
https://ui.adsabs.harvard.edu/abs/1991A&A...248..669M

Mein P., *The MSDP of THEMIS: capabilities, first results and prospects*, 2002, Astron. Astrophys., 381, 271
https://ui.adsabs.harvard.edu/abs/2002A&A...381..271M

Mein, P., Mein, N., Bommier, V., *"Fast imaging spectroscopy with MSDP spectrometers: Vector magnetic maps with THEMIS/MSDP"*, 2009, Astron. Astrophys., 507(1), 531.
https://ui.adsabs.harvard.edu/abs/2009A&A...507..531M

Mein, P., Malherbe, J.M., Sayède, F., Rudawy, P., Phillips, K., Keenan, F., 2021, *"Four decades of advances from MSDP to S4I and SLED imaging spectrometers"*, Solar Phys., 296, 30.

Rompolt, B., Mein, P., Mein, N., Rudawy, P., Berlicki, A., *"The MSDP Recently Installed at the Spectrograph of the Large Coronograph"*, 1994, JOSO Annual Report, 93, 95

Roudier, Th., Malherbe, J.-M., Moity, J., Rondi, S., Mein, P., Coutard, Ch., 2006. *"Sub arcsec evolution of solar magnetic fields"*, Astron. Astrophys., 455, 1091.

Sayède, F., January 2023, EST Praha meeting.

Semel, M., 1980, ''*Un analyseur précis de polarisation optique*'', Astron. Astrophys., 91, 369.